\documentclass[aps,prl,groupedaddress]{revtex4}
\usepackage{graphicx}
\begin{document}

%\preprint{}

[hep-ph/0301069] \hfill Saclay/SPhT--T02/183

\title{All-Orders Singular Emission in Gauge Theories}


\author{David A. Kosower}
\email[]{kosower@spht.saclay.cea.fr}
%\thanks{}
\affiliation{
Service de Physique Th\'eorique\footnote{Laboratory of the
{\it Direction des Sciences de la Mati\`ere\/}
of the {\it Commissariat \`a l'Energie Atomique\/} of France.}\\
Centre d'Etudes de Saclay\\
F--91191 Gif-sur-Yvette cedex\\
France}

\date{\today}

\begin{abstract}
I present a class of functions unifying all singular limits for the
emission of soft or collinear gluons in gauge-theory amplitudes at any
order in perturbation theory.  Each function is a generalization of
the antenna functions of
ref.~\cite{SingleAntenna,MultipleAntenna}.  The helicity-summed
interferences these functions are thereby also generalizations to
higher orders of the Catani--Seymour dipole factorization function.
\end{abstract}

\pacs{}
%\keywords{}

\maketitle

\section{}
%\subsection{}
%\subsubsection{}

\def\tp{\!+\!}\def\tm{\!-\!}
\def\tc{\!\cdot\!}
\def\Ant{\mathop{\rm Ant}\nolimits}
\def\pol{\varepsilon}
\def\e{\epsilon}
\def\tree{{\rm tree\vphantom{p}}}
\def\LIPS{{\rm LIPS}}
\def\phpol{{\rm ph.\ pol.}}
\def\oneloop{{1\lloop}}
\def\lloop{\dash{}{\rm loop}}
\def\dash{\hbox{-\kern-.02em}}
\def\llongrightarrow{%
\relbar\mskip-0.5mu\joinrel\mskip-0.5mu\relbar\mskip-0.5mu\joinrel\longrightarrow}
\def\inlimit^#1{\buildrel#1\over\llongrightarrow}
Amplitudes in gauge theories have universal factorization and scaling
behaviors as sets of massless momenta become collinear or soft.  The
study of these factorization properties goes back to the earliest
quantum-mechanical studies of soft-photon emission by Bloch and
Nordsieck~\cite{BlochNordsieck}.  It has played an important role in
our ability to make increasingly accurate predictions for scattering
processes at high-energy colliders.  An understanding of the
factorization properties are necessary both to predictions at fixed order,
 and those relying on a summation of dominant logarithms.

Recent progress in two-loop calculations~\cite{TwoloopSummary} has
opened the way for next-to-next-to-leading order (NNLO) calculations
of jet production, both at lepton and hadron colliders.  Completing
this program, and obtaining numerical programs, will require further
work on integrals over singular regions of gluon and quark-pair
emission.  These integrals will be rendered more tractable by a
formalism which unifies the factorization behavior of amplitudes in
the disparate collinear, soft, or mixed regions of phase space.
Catani and Seymour proposed~\cite{CataniSeymour} such a formalism, the
so-called dipole formalism, for one singular emission (one collinear
pair or one soft gluon).  I later wrote down~\cite{SingleAntenna} an
equivalent formalism, at the level of the amplitude rather than the
squared matrix element.  This formalism
generalizes~\cite{MultipleAntenna} to the emission of an arbitrary
number of singular partons in tree-level amplitudes.
The integrals
over the factorization functions as further computed using a dimensional
regulator in ref.~\cite{CataniSeymour} summarize in a universal fashion
the infrared poles required to cancel
those in one-loop virtual corrections.

\def\ah{{\hat a}}
\def\bh{{\smash{{\hat b}}{}}}
\def\phpol{{\rm ph.\ pol.}}
\def\back{\hskip -15mm}
Define an {\it antenna\/} function or amplitude via
\begin{equation}
\Ant(\ah,\bh\leftarrow a,1,\ldots,m,b) =
\sum_{j=0}^m J(a,1,\ldots,j;\ah)
             J(j\tp1,\ldots,m,b;\bh),
\label{MultipleEmissionAntennaDef}
\end{equation}
where $J$ is a gluon (or quark) current as used in the Berends--Giele
recurrence relations~\cite{Recurrence}.  
In the form given by Dixon~\cite{DixonTASI},
the gluon current (with opposite sign to Dixon's) is,
\def\spacer{ \hphantom{ -{i d_{\mu\mu'}(K_{1,n}) \over K_{1,n}^2}\Biggl[} }
\def\back{\hskip -15mm}
\def\L{\left(}\def\R{\right)}
\def\LB{\left[}\def\RB{\right]}
\def\LP{\left.}\def\RP{\right.}
\def\RV{\right|}
\begin{eqnarray}
J_\mu(1,\ldots,n) =&&
\nonumber\\ &&\back -{i d_{\mu\mu'}(K_{1,n}) \over K_{1,n}^2}
\LB \sum_{j=1}^{n-1} 
V_3^{\mu'\nu'\rho'}(K_{1,j},K_{j\tp1,n},-K_{1,n})
d_{\nu\nu'}(K_{1,j}) d_{\rho\rho'}(K_{j\tp1,n}) 
%\RP\nonumber\\ &&\back\spacer\times\LP
J^\nu(1,\ldots,j)
J^\rho(j\tp1,\ldots,n)
\RP\nonumber\\ &&\back\spacer\LP
-\sum_{j_1=1}^{n-1}\sum_{j_2=j_1+1}^{n-1}
V_4^{\mu'\nu'\rho'\lambda'}
d_{\nu\nu'}(K_{1,j_1}) d_{\rho\rho'}(K_{j_1\tp1,j_2}) 
   d_{\rho\rho'}(K_{j_2\tp1,n})
\RP\\ &&\spacer\hphantom{\sum\sum\sum}\times\LP\vphantom{\sum_{j=1}^1}
J^\nu(1,\ldots,j_1)
J^\rho(j_1\tp1,\ldots,j_2)
J^\lambda(j_2\tp1,\ldots,n)
\RB.\nonumber
\label{RecurrenceRelations}
\end{eqnarray}
Here, $K_{i,j} = k_i + \cdots + k_j$; in 
eqn.~(\ref{MultipleEmissionAntennaDef}),
$J(1,\ldots,n;x) = \pol_x\cdot J(1,\ldots,n)$ and 
the currents are to be evaluated in light-cone gauge, for which
\begin{eqnarray}
&&V_{3}^{\mu\nu\rho}(P,Q,R) = {i\over\sqrt2}
\LB g^{\nu\rho} (P\tm Q)^\mu \tp g^{\rho\mu} (Q\tm R)^\nu 
    \tp g^{\mu\nu} (R\tm P)^\rho\RB;
\nonumber\\ &&
V_4^{\mu\nu\rho\lambda} = {i\over2}\LB 2 g^{\mu\rho} g^{\nu\lambda}
\tm g^{\mu\nu} g^{\rho\lambda} \tm g^{\mu\lambda} g^{\nu\rho}\RB;
\hskip 5mm d_{\mu\nu}(k) = -g_{\mu\nu} + {q^\mu k^\nu \tp k^\mu q^\nu\over q\cdot k}.
\label{LightConeGauge}
\end{eqnarray}

The antenna amplitude describes in a unified way all leading
singularities of tree
amplitudes as the color-connected set of momenta $\{k_a,k_1,\ldots,k_i\}$ 
becomes
collinear, likewise for $\{k_j,\ldots,k_n,k_b\}$, and as the momenta
$\{k_{i\tp1},\ldots,k_{j\tm1}\}$ become soft,
%\begin{eqnarray}
\begin{equation}
A_n(\ldots,a,1,\ldots,m,b,\ldots) 
\inlimit^{k_1,\ldots,k_m {\rm\ singular}} %&&
%\nonumber\\ &&\back
\mskip -8mu\sum_{\phpol\ \lambda_{a,b}}
\Ant(\ah^{\lambda_a},\bh^{\lambda_b}\leftarrow a,1,\ldots,m,b)
A_{n-m}(\ldots,-k_{\ah}^{-\lambda_a},-k_{\bh}^{-\lambda_b},\ldots) .
\nonumber
\label{TreeEmissionFactorization}
\end{equation}
%\end{eqnarray}
The momenta $k_{\ah,\bh}$ are reconstructed from the original momenta via
{\it reconstruction\/} functions given in ref.~\cite{MultipleAntenna}.

In this Letter, I generalize the construction of
ref.~\cite{MultipleAntenna} to higher orders in perturbation theory.
To obtain such a generalization, we must first write down a formula
for the higher-loop analog of the current $J$.  (See
ref.~\cite{CataniGrazziniSoft} for a related construction.)  I will
restrict attention here to leading-color amplitudes in the context of a
color decomposition~\cite{Color}, so that only planar diagrams need be
considered.  These higher-loop analogs to the current will bear the
same relation to higher-loop splitting amplitudes as do the tree-level
currents to the tree-level multi-collinear splitting amplitudes:
spinor products replace momentum fractions, adding phase and
correlation information, and capturing a larger scope of singular
behavior.

\def\ll#1{{\lambda_{#1}}}
Higher-loop currents $J^{l\lloop}$ may be defined via their cuts,
\begin{eqnarray}
&&\LP J^{l\lloop}(1^\ll1,2^\ll2,\ldots,m^\ll{m};P)
   \RV_{t_{c\cdots d} {\rm\ cut}} =
\nonumber\\ &&\hskip 5mm\sum_{k=0}^{l-1} 
\,\sum_{j=2}^{l+ 1- k} 
\sum_{\phpol\ \sigma_i}\int d\LIPS^{4-2\e}(\ell_1,\ldots,\ell_j)\;
%\nonumber\\ &&\hskip 25mm\times
 J^{k\lloop}(1^\ll1,\ldots,(c\tm1)^\ll{c\tm1},
             \ell_1^{-\sigma_1},\ldots,\ell_j^{-\sigma_j},(d\tp1)^\ll{d\tp1},
             \ldots,m;P)
\nonumber\\ &&\hskip 75mm\times
   \,A_{d\tm c\tp j\tp1}^{(l\tp 1\tm j\tm k)\lloop}
  (c^\ll{c},\ldots,d^\ll{d},-\ell_j^{\sigma_j},\ldots,-\ell_1^{\sigma_1}).
%\nonumber
%
\label{HigherOrderJ}
\end{eqnarray}
In this equation, $X^{0\lloop}$ means $X^{\tree}$.
While the currents appearing here must be evaluated in light-cone
gauge, the on-shell amplitudes on the other side of the cut may be
evaluated in any gauge.

\begin{figure}
 \includegraphics{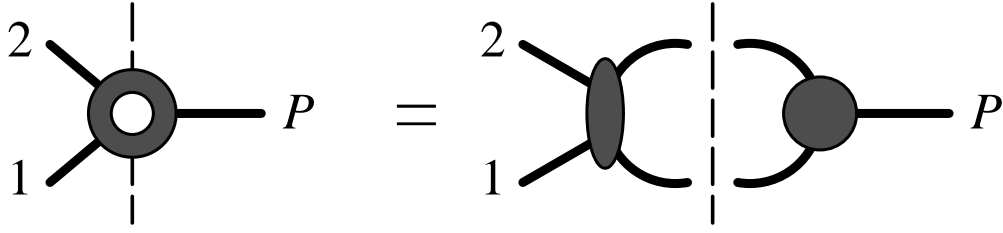}%
 \caption{\label{ThreePointFigure}%
The one-loop three-point current is defined by its cut.}
\end{figure}

In the one-loop case, the three-point current has only one cut, illustrated
in fig.~1, and we can reconstruct a loop integral from it,
\begin{eqnarray}
J(1,2;P) = &&
%\nonumber\\ &&\back
\mskip -25mu\sum_{\phpol\ \sigma_1,\sigma_2} 
\int {d^{4-2\e}\ell\over (2\pi)^{4-2\e}}\;{i\over\ell^2}
 J^{\tree}((\ell\tp a\tp b)^{-\sigma_2},
-\ell^{-\sigma_1};P){i\over (\ell\tp k_a\tp k_b)^2}
%\nonumber\\ &&\back\hskip 40mm\times
      A_{4}^\tree((-\ell\tm a\tm b)^{\sigma_2},1,2,\ell^{\sigma_1}).
%\nonumber
\label{OneLoopJ}
\end{eqnarray}
The restriction to physical polarizations is important, as it will give
rise to projection operators inside the loop.
More generally, eqn.~(\ref{HigherOrderJ}) gives the absorptive part of
the higher-loop current.  The dispersive part may in principle be obtained
through a dispersion relation in $D=4-2\e$ dimensions (where no subtractions
are needed~\cite{vanNeerven,Unitarity}).  In practice, the reconstruction
of loop integrals from combining different cuts is probably an easier
way to proceed.  The computation of the three-point one-loop current 
is very similar to that of the one-loop splitting 
amplitude~\cite{OneLoopSplitting}, 
and one obtains for the unrenormalized current,
\def\cg{c_\Gamma}
\begin{eqnarray}
&& J(1,2;P) =
-{\cg} \L{\mu^2\over -s_{12}}\R^{\e}
   \LB F_0({q\cdot k_1\over q\cdot(k_1+k_2)}) 
       +F_0({q\cdot k_2\over q\cdot(k_1+k_2)}) \RB\,J^\tree(1,2;P) 
\nonumber\\ && \hphantom{J(1,2;P) =} - {1\over\sqrt2 s_{12}^2}
 {\cg (1-\e\delta)\over (1-2\e)(1-\e)(3-2\e)}\L{\mu^2\over -s_{12}}\R^{\e}
         \,(k_1-k_2)\cdot\pol_P\, 
       \L s_{12} \pol_1\cdot\pol_2 - 2 k_2\cdot \pol_1\,k_1\cdot \pol_2\R.
\end{eqnarray}
\def\Ord{{\cal O}}
\def\F#1#2{\,{{\vphantom{F}}_{#1}F_{#2}}}
\def\Li{\mathop{\rm Li}\nolimits}
The parameter $\delta$ determines the variant of dimensional regularization
used, $\delta = 0$ for the four-dimensional helicity scheme~\cite{FDH},
and $\delta = 1$ for the conventional scheme~\cite{CollinsBook}.
In this equation, $\cg = {\Gamma(1+\epsilon)\Gamma^2(1-\epsilon)\over
  (4\pi)^{2-\epsilon}\Gamma(1-2\epsilon)}$,
and (with $\F21$ the Gauss hypergeometric function and $\Li_2$
the dilogarithm),
\begin{eqnarray}
&&F_0(w) = 
%%%%% begin F0 : F0p[w_]
            {1\over\e^2}\LB{\Gamma(1-\e)\Gamma(1+\e)}
                            \,w^{-\e} (1-w)^\e
                        +{1\over2}
                        -(1-w)^\e \,\F21\L \e,\e;1+\e;w\R\RB\cr
%%%%% end F0 : F0p[w_]
\nonumber\\ &&\hphantom{F_0(w) }
%%%%% begin F0 : F0pexpand[w_]
= {1\over\e^2}\LB{\Gamma(1-\e)\Gamma(1+\e)}
                            \,w^{-\e} (1-w)^\e
                        +{1\over2}
                        -(1-w)^\e
                         -\e^2 \Li_2(w) \RB + \Ord(\e).
%%%%% end F0 : F0pexpand[w_]
\end{eqnarray}

With the higher-order current $J^{l\lloop}$ in hand, we can write down an
expression for the higher-loop generalization of the antenna amplitude,
\begin{equation}
\Ant^{n\lloop}(\ah,\bh\leftarrow a,1,\ldots,m,b) =
%\cr &\hskip 10mm
\sum_{j=0}^m \sum_{l=0}^n J^{l\lloop}(a,1,\ldots,j;\ah)
             J^{(n-l)\lloop}(j\tp1,\ldots,m,b;\bh).
\label{HigherLoopAntennaDef}
\end{equation}

We can derive the factorization of the leading-color~\cite{Color}
contribution to higher-loop amplitudes by matching on to
known purely-collinear limits~\cite{AllOrdersCollinear}.  We then find
for the corresponding factorization,
\begin{eqnarray}
&&A_n^{r\lloop}(\ldots,a,1,\ldots,m,b,\ldots) 
\inlimit^{k_1,\ldots,k_m {\rm\ singular}}
\nonumber\\ &&\hskip 15mm
\sum_{\phpol\ \lambda_{a,b}} \sum_{v=0}^r 
  \Ant^{v\lloop}(\ah^{\lambda_a},\bh^{\lambda_b}\leftarrow a,1,\ldots,m,b) 
   A_{n-m}^{(r-v)\lloop}(\ldots,-k_\ah^{-\lambda_a},
                 -k_\bh^{-\lambda_b},\ldots).
\label{MultipleEmissionAntennaFactorization}
\end{eqnarray}
Multi-collinear limits in $m$ momenta arise from the simultaneous
vanishing of invariants in those momenta and one of the two hard
momenta $a$ or $b$.  Mixed collinear-soft (or pure
multi-soft) singularities arise from the vanishing of additional
invariants involving the other hard momentum as well.  
The triply-collinear limit $k_a\parallel k_1 \parallel k_2$, for example,
arises when $t_{a12}$, $s_{a1}$, and $s_{12}$ all vanish at a similar
rate.  A mixed limit, for example $k_2$ becoming soft, is reflected in
the vanishing of additional invariants, in this particular case $s_{2b}$.

Because the leading singular behavior in
such additional invariants is already included in the antenna
amplitude, it also captures the leading behavior in these mixed limits.
(This is already implicit in collinear splitting amplitudes, which
have the correct $z\rightarrow 0$ behavior to describe soft regions,
but lack the phase information required for a complete description in
those regions.)  Accordingly,
eqn.~(\ref{MultipleEmissionAntennaFactorization}) gives the leading
behavior of $r\lloop$ leading-color amplitudes in all singular limits
involving the color-connected momenta $k_1,\ldots,k_m$.  The singular
behavior of leading-color amplitudes
in limits of color-nonconnected sets of momenta can be built
up from products of antenna functions.

The one-loop single-emission case antenna amplitude
was considered previously by Uwer and 
the author~\cite{OneLoopSplitting}.  Freely
adding terms less singular than the leading ones in all limits,
and judiciously multiplying collections
of terms less
singular than $1/E_1$ in the soft limit $k_1\rightarrow 0$ 
in the current $J^\oneloop(a,1;\ah)$ by $(s_{1b}+s_{1a})^\e s_{1b}^{-\e}$
(a factor which is one in the collinear limit $k_1\parallel k_a$), 
and similarly
for the current $J^\oneloop(1,b;\bh)$, one obtains,
\def\dAnt{\Ant^{F}}
\begin{eqnarray}
&& \Ant^\oneloop(\ah,\bh\leftarrow a,1,b) =
%%%%% begin Ant : Ant1loop
-{\cg} \L{\mu^2 K^2\over -s_{a1}s_{1b}}\R^{\e}
   \biggl\{ {\Gamma(1-\e)\Gamma(1+\e) \over\e^2}
        \biggl[2 - \biggl({s_{ab}\over K^2}\biggr)^\e\biggr]
       + F\L {s_{ab}\over K^2}\R\biggr\}\,
%%%%% end Ant : Ant1loop
%\nonumber\\ && \hphantom{ \Ant^\oneloop(\ah,\bh\leftarrow a,1,b) =}\times 
 \Ant^\tree(\ah,\bh\leftarrow a,1,b)
\nonumber\\ && \hphantom{ \Ant^\oneloop(\ah,\bh\leftarrow a,1,b) = }
%%%%% begin Ant : AntFcoeff
-{\cg(1-\e \delta)\over (1-2\e)(1-\e)(3-2\e)}
\biggl({\mu^2 K^2\over -s_{a1} s_{1b}}\biggr)^{\e} 
\biggl( 1-{s_{ab}\over K^2}\biggr)^\e
%%%%% end Ant : AntFcoeff
\dAnt(\ah,\bh\leftarrow a,1,b)
%\nonumber
\end{eqnarray}
where $K= k_a+k_1+k_b$,
\begin{eqnarray}
&& F(w) = 
%%%%% begin F : Fp[w_]
            {1\over\e^2} \bigl[ \Gamma(1-\e)\Gamma(1+\e) 
                           (w^{-\e} (1-w)^{2\e}+ 2 w^\e -2)
                           +(1-w)^\e 
                           - (1-w)^{2\e}\F21(\e,\e;1+\e;w)
\nonumber\\ && \hphantom{ F(w) = {1\over\e^2} \bigl[ }
                            - w^\e (1-w)^\e \F21(\e,\e;1+\e;1-w)\bigr]
%%%%% end F : Fp[w_]
\\ && \hphantom{F(w) } 
%%%%% begin F : Fpexpand[w_]
= \ln w \ln \bigl[ w/(1-w)^2\bigr]+\Ord(\e),
%%%%% end F : Fpexpand[w_]
\nonumber
\end{eqnarray}
and
\begin{eqnarray}
&&\dAnt(\ah,\bh\leftarrow a,1,b) = 
         L(a,1,\ah;b,\bh) + L(1,b,\bh;a,\ah),
\\ && L(p,q,r;u,v) =
{1\over\sqrt2 s_{pq}^2}(k_p-k_q)\cdot\pol_{r}\, \pol_u\cdot\pol_{v}
       \bigl( s_{pq} {\pol_p\cdot\pol_q} 
          - {2 k_q\cdot \pol_p\,k_p\cdot \pol_q}\bigr).
\end{eqnarray}

\def\spa#1.#2{\left\langle#1\,#2\right\rangle}
\def\spb#1.#2{\left[#1\,#2\right]}
\def\sepr{\quad\!}
The new helicity structure has non-vanishing values for the following
helicity configurations,
\begin{eqnarray}
&& \dAnt(\ah^+,\bh^+\leftarrow a^+,1^+,b^-) = 
%%%%% begin AntF : AntF[PPPPM]
-{\spa{a}.b\spb{a}.1\spa1.{b}\over \spa{\ah}.{\bh}^2 \spa{a}.1^2},
%%%%% end AntF : AntF[PPPPM]
\sepr
\dAnt(\ah^+,\bh^+\leftarrow a^-,1^+,b^+) = 
%%%%% begin AntF : AntF[PPMPP]
-{\spa{a}.b\spa{a}.1\spb1.{b}\over \spa{\ah}.{\bh}^2 \spa1.{b}^2},
%%%%% end AntF : AntF[PPMPP]
\sepr
\nonumber\\ && 
\dAnt(\ah^+,\bh^+\leftarrow a^-,1^-,b^-) = 
%%%%% begin AntF : AntF[PPMMM]
{\spa{a}.b (s_{a1} + s_{1b})\over \spa{\ah}.{\bh}^2 \spb{a}.1\spb1.{b}},
%%%%% end AntF : AntF[PPMMM]
\sepr
\dAnt(\ah^+,\bh^-\leftarrow a^+,1^+,b^+) = 
%%%%% begin AntF : AntF[PMPPP]
-{\spa{a}.{\bh}^2 \spb{\ah}.b^2 \spb{a}.1 \spa1.{b}\over
  \spa{a}.{b} \spa{\ah}.{\bh}^2 \spb{\ah}.{\bh}^2 \spa{a}.1^2},
%%%%% end AntF : AntF[PMPPP]
\sepr
\nonumber\\ && 
\dAnt(\ah^+,\bh^-\leftarrow a^-,1^+,b^+) = 
%%%%% begin AntF : AntF[PMMPP]
{\spa{a}.{\bh}^2 \spb{\ah}.b^2 \spb{a}.1 \over
   \spa{\ah}.{\bh}^2 \spb{a}.b \spb{\ah}.{\bh}^2\spa1.{b}},
%%%%% end AntF : AntF[PMMPP]
\sepr
\dAnt(\ah^+,\bh^-\leftarrow a^-,1^-,b^+) = 
%%%%% begin AntF : AntF[PMMMP]
-{\spa{a}.{\bh}^2 \spb{\ah}.{b}^2\spa1.{b} \over
  \spa{a}.{b} \spa{\ah}.{\bh}^2 \spb{\ah}.{\bh}^2 \spb{a}.1 },
%%%%% end AntF : AntF[PMMMP]
\sepr
\nonumber\\ && 
\dAnt(\ah^+,\bh^-\leftarrow a^-,1^-,b^-) = 
%%%%% begin AntF : AntF[PMMMM]
 {\spa{a}.{\bh}^2 \spb{\ah}.b^2\spb{a}.1 \spa1.{b} \over
   \spa{\ah}.{\bh}^2 \spb{a}.b \spb{\ah}.{\bh}^2 \spb1.{b}^2}.
%%%%% end AntF : AntF[PMMMM]
\label{NonvanishingHelicities}
\end{eqnarray}
including a number of configurations for which the tree-level antenna
amplitude vanishes.   It vanishes for,
%the helicity configurations,
\begin{eqnarray}
&& \ah^+,\bh^+\leftarrow a^+,1^+,b^+;\sepr
\ah^+,\bh^+\leftarrow a^+,1^-,b^+;\sepr
\ah^+,\bh^+\leftarrow a^-,1^+,b^-;\sepr
\ah^+,\bh^+\leftarrow a^+,1^-,b^-;\sepr
\ah^+,\bh^+\leftarrow a^-,1^-,b^+;\sepr
\nonumber\\ &&
\ah^+,\bh^-\leftarrow a^+,1^+,b^-;\sepr
\ah^+,\bh^-\leftarrow a^+,1^-,b^-;\sepr
\ah^+,\bh^-\leftarrow a^+,1^-,b^+;\sepr
{\rm and\ }\ah^+,\bh^-\leftarrow a^-,1^+,b^-.\sepr
\label{VanishingHelicities}
\end{eqnarray}
The remaining helicity configurations 
%besides those
%listed explicitly in eqns.~(\ref{VanishingHelicities},%
%\ref{NonvanishingHelicities})
can be obtained via parity.

In calculations of higher-order corrections to differential matrix
elements, infrared singularities arise from two sources.  These are
the virtual corrections on the one hand, and integrals over soft or
collinear phase space on the other.  The singularities arising in 
the latter are captured in their entirety in integrals over singular
phase space of interferences of various antenna amplitudes.  At
next-to-leading order, the relevant integral is that of the
tree-level one-emission antenna amplitude squared, 
$|\Ant^tree(\ah,\bh\leftarrow a,1,b)|^2$, for which an expression
was given in refs.~\cite{SingleAntenna,MultipleAntenna}.  At 
next-to-next-to-leading order, two integrals are required,
one being that of the tree-level double-emission antenna amplitude squared,
$|\Ant^\tree(\ah,\bh\leftarrow a,1,2,b)|^2$, given in 
ref.~\cite{MultipleAntenna}.  The other required integral is that
of the one-loop--tree interference (summed over the helicities of legs $a$,
$1$, and $b$, and averaged over those of $\ah$ and $\bh$),
\begin{eqnarray}
&& 2 \mathop{\rm Re}\bigl[ \Ant^{\oneloop*}(\ah,\bh\leftarrow a,1,b)
            \Ant^{\tree}(\ah,\bh\leftarrow a,1,b)
            \bigr] =
\nonumber\\ && \hskip 15mm
%%%%% begin Interference : interference
-4\cg \L{\mu^2 K^2\over -s_{a1}s_{1b}}\R^{\e}
   \biggl\{ {\Gamma(1-\e)\Gamma(1+\e) \over\e^2}
        \biggl[2 - \biggl({s_{ab}\over K^2}\biggr)^\e\biggr]
       + F\L {s_{ab}\over K^2}\R\biggr\}\,
{\left( K^2 (s_{a1}+s_{1b}) + s_{ab}^2\right)^2
           \over s_{a1} s_{1b} s_{ab} (K^2)^2}
\\ && \hskip 15mm
-{2\cg\over (1-2\e)(1-\e)(3-2\e)}
   \biggl({\mu^2 K^2\over -s_{a1} s_{1b}}\biggr)^{\e} 
\biggl( 1-{s_{ab}\over K^2}\biggr)^\e
   \biggl\{ {1\over s_{a1}}+{1\over s_{1b}}
            -\e \delta {s_{ab}\over (K^2)^2} \Bigl(
               {s_{a1}\over s_{1b}}+{s_{1b}\over s_{a1}}\Bigr)
   \biggr\}
%%%%% end Interference : interference
\nonumber
\end{eqnarray}

I thank Z.~Bern for helpful comments.


\end{document}